%% file: note.tex
\documentclass[a4paper,10pt]{article}

\usepackage{jheppub} 

\usepackage[T1]{fontenc} 

    \title{Precise predictions for boosted Higgs production}

    \abstract{Inclusive Higgs boson production at large transverse
      momentum is induced by different production channels. We focus
      on the leading production mechanism through gluon fusion, and
      perform a consistent combination of the state of the art
      calculations obtained in the infinite-top-mass effective theory
      at next-to-next-to-leading order (NNLO) and in the full Standard
      Model (SM) at next-to-leading order (NLO). We thus present
      approximate QCD predictions for this process at NNLO, and a
      study of the corresponding perturbative uncertainties.
      This calculation is then compared with those obtained with
      commonly used event generators, and we observe that the
      description of the considered kinematic regime provided by these
      tools is in good agreement with state of the art calculations.
      Finally, we present accurate predictions for other production
      channels such as vector boson fusion, and associated production
      with a gauge boson, and with a $t\bar{t}$ pair. We find that, at
      large transverse momentum, the contribution of other production
      modes is substantial, and therefore must be included for a
      precise theory prediction of this observable.}

\usepackage{graphicx}

\usepackage{caption}
\usepackage{subcaption}
\usepackage{lineno}
\usepackage{changepage}

\author[]{Conveners of the gluon-fusion Working Group: \\}
\author[a]{K. Becker,}
\author[b]{F. Caola,}
\author[c]{A. Massironi,}
\author[d]{B. Mistlberger,}
\author[e]{P. F. Monni.}
\author[]{\\}
\author[]{\\In collaboration with:\\}
\author[f,g]{X. Chen,}
\author[h]{S. Frixione,}
\author[g]{T. Gehrmann,}
\author[i]{N. Glover,}
\author[j]{K. Hamilton,}
\author[e]{A. Huss,}
\author[i]{S. P. Jones,}
\author[e]{A. Karlberg,}
\author[k]{M. Kerner,}
\author[l]{K. Kudashkin,}
\author[m]{J. M. Lindert,}
\author[n]{G. Luisoni,}
\author[e]{M. L. Mangano,}
\author[g]{S. Pozzorini,}
\author[o,p]{E. Re,}
\author[b,q]{G. P. Salam,}
\author[r]{E. Vryonidou,}
\author[s]{C. Wever.}

\affiliation[a]{Department of Physics, University of Warwick,
  Coventry, CV4 7AL, UK}
\affiliation[b]{Rudolf Peierls Centre for Theoretical Physics,Oxford 
  University, OX1 3PU, UK}
\affiliation[c]{INFN, Sezione di Milano-Bicocca, Piazza della Scienza 3, 20126 Milano, Italy}
\affiliation[d]{SLAC National Accelerator Laboratory, Stanford University, Stanford, CA 94039,
USA}
\affiliation[e]{CERN, Theoretical Physics Department, CH-1211 Geneva 23, Switzerland}
\affiliation[f]{School of Physics, Shandong University, Jinan, Shandong 250100, China}
\affiliation[g]{Physik-Institut, Universit\"at Z\"urich, Winterthurerstrasse 190, CH-8057 Z\"urich, Switzerland}
\affiliation[h]{INFN, Sezione di Genova, Via Dodecaneso 33, I-16146,  Genoa, Italy}
\affiliation[i]{Institute for Particle Physics Phenomenology,   Department of Physics, University of Durham, Durham, DH1 3LE, UK}
\affiliation[j]{Department of Physics and Astronomy, University College London, London, WC1E 6BT, UK}
\affiliation[k]{Institute for Theoretical Physics, Karlsruhe Institute of Technology, 76128 Karlsruhe, Germany}
\affiliation[l]{Tif Lab, Dipartimento di Fisica, Università di Milano and
INFN, Sezione di Milano, Via Celoria 16, I-20133 Milano, Italy}
\affiliation[m]{Department of Physics and Astronomy, University of Sussex, Brighton BN1 9QH, UK}
\affiliation[n]{Max-Planck-Institut f\"ur Physik, F\"ohringer Ring 6,
  80805 M\"unchen, Germany}
\affiliation[o]{Dipartimento di Fisica G. Occhialini,
Università degli Studi di Milano-Bicocca and INFN, Sezione di Milano-Bicocca}
\affiliation[p]{LAPTh, Universit\'e Grenoble Alpes, Universit\'e Savoie Mont Blanc, CNRS, 74940 Annecy, France}
\affiliation[q]{All Souls College, Oxford OX1 4AL, UK}
\affiliation[r]{Department of Physics and Astronomy, University of Manchester,
Oxford Road, Manchester M13 9PL, United Kingdom}
\affiliation[s]{Physik-Department T31, Technische Universit\"at M\"unchen,
James-Franck-Strasse 1, D-85748 Garching, Germany}

\newcommand{\mtop}{m_t}

\preprint{LHCHXSWG-2019-002, CERN-TH-2020-074, TUM-HEP-1264/20}

\begin{document} 
\maketitle
\flushbottom

\section{Introduction}

The continuously increasing amount of data recorded at the LHC opens
the possibility to explore properties of the Higgs boson in a
multitude of kinematic regimes.  Of particular interest is the
transverse momentum distribution of the Higgs boson for very large
transverse momenta.  Measurements of this observable allow for unique
insights into the microscopic structure of the interactions of the
Higgs boson with strongly interacting particles and might shed light
on physics beyond the Standard Model.  The observation of the Higgs
boson in this kinematic regime is however extremely challenging.

The inclusive search for the Standard Model Higgs boson produced at
large transverse momentum ($p_\perp$), and decaying to a bottom
quark-antiquark pair, has been performed using data collected in pp
collisions at $\sqrt{s}=13$ TeV by the CMS and ATLAS
experiments~\cite{Sirunyan:2017dgc,Aad:2019uoz,CMS:2020zge,ATLAS:2021tbi,ATLAS:2022fnp,CMS:2022wpo}.

It is the objective of this document to study accurate theoretical
predictions for the transverse momentum distribution with
$p_\perp > 400$ GeV.  We present new, state of the art predictions for
the dominant gluon-fusion induced production of a Higgs boson and at
least one hard partonic jet that recoils against it, based on
perturbative QCD computations at a centre-of-mass energy of $13$ TeV.
In particular we perform a combination of next-to-next-to leading
order (NNLO) calculations in the heavy top quark effective
theory~\cite{Boughezal:2015aha,Boughezal:2015dra,Caola:2015wna,Chen:2016zka,Campbell:2019gmd,Chen:2019wxf,Chen:2021ibm}
with next-to-leading order (NLO) predictions in the full SM with
finite top-quark mass
($\mtop$)~\cite{Lindert:2018iug,Jones:2018hbb,Bonciani:2022jmb}. A
related combination has been recently presented in
ref.~\cite{Chen:2021azt}. We provide a recommendation for the
theoretical prediction for the gluon-fusion channel to be used by the
ATLAS and CMS collaborations. Subsequently, we compare these
predictions with state of the art hard-event
generators~\cite{Alioli:2008tz,Campbell:2012am,Hamilton:2012rf,Frederix:2016cnl,Hamilton:2013fea}.
We find that indeed the most advanced event generators describe the
cross sections of interest within uncertainties. Furthermore, we also
report the contributions from the vector boson fusion, VH, and
${\rm t\bar t H}$ production modes for the observable under
consideration, together with a NLO calculation of the electro-weak
corrections.

\section{Predictions for the gluon-fusion channel}
\label{sec:predictions}
We start by focusing on the predictions for the gluon-fusion (ggF)
channel, and by giving an approximate NNLO result, which we quote as
our recommendation for the cross section in the boosted regime. This
is obtained by combining the following two predictions for the
production of a Higgs boson and at least one partonic jet: the NNLO
(${\cal O}(\alpha_s^5)$) calculation in the large-$\mtop$ limit and
the NLO (${\cal O}(\alpha_s^4)$) calculation in the full SM.

The setup used for the NNLO results in the large-$\mtop$ limit is as
follows
\begin{itemize}
\item pp collisions at $\sqrt{s} = 13$ TeV,
\item $m_H=125$~GeV, $m_t=173.2$~GeV, all other parameters as per
  YR4~\cite{YR4},
\item {\tt PDF4LHC15\_nnlo\_mc},
\item central scales $\mu_F=\mu_R= M_{T,H}$, where we defined the
  Higgs transverse mass
\begin{equation}
\label{eq:MT}
M_{T,H}=\sqrt{m_H^2
  + p_{\perp}^2}\,.
\end{equation}
\item In our predicions we consider an on-shell Higgs boson, so we do
  not include any particular decay.
\end{itemize}

In Section~\ref{sec:generators}, we also consider the predictions from
common event generators. Such predictions come with their own scale
setting, as reported in the discussion below. The above scale choice
is of course not unique, and different choices lead to differences in
the final predictions. However, the goal of this manuscript is to
compare different theory predictions for the observable under
study. Therefore, we limit ourselves to the above choice for the
discussion that follows.

\input{Chapters/FO.tex}

\input{Chapters/MC.tex}

\input{Chapters/Channels.tex}

\input{Chapters/Conclusions.tex}

\appendix

\input{Chapters/gluon_fusion_high_pt.tex}

\bibliography{biblio}
\bibliographystyle{JHEP}

\end{document}

%% file: Chapters/FO.tex
\subsection{Fixed-order}
\label{sec:fixed}
In this section we present state of the art predictions for the
transverse momentum ($p_\perp$) spectrum of the Higgs boson in the
boosted regime.  The transverse momentum distribution was computed at
NNLO in perturbative QCD in the heavy top quark effective theory (EFT)
in
refs.~\cite{Boughezal:2015aha,Boughezal:2015dra,Chen:2016zka,Campbell:2019gmd}.
Specifically,
refs.~\cite{Boughezal:2015aha,Boughezal:2015dra,Chen:2016zka,Campbell:2019gmd}
compute NNLO corrections to the Born level production of a Higgs boson
and a jet.  In the EFT approximation the top quark is treated as
infinitely heavy and its degrees of freedom are integrated out.  It is
however well known that the pure EFT computation fails to describe the
$p_\perp$ spectrum for transverse momenta larger than $\sim200$
GeV, cf.~\cite{Jones:2018hbb}.

\begin{figure*}[htb]
\centering
\includegraphics[width=.65\textwidth]{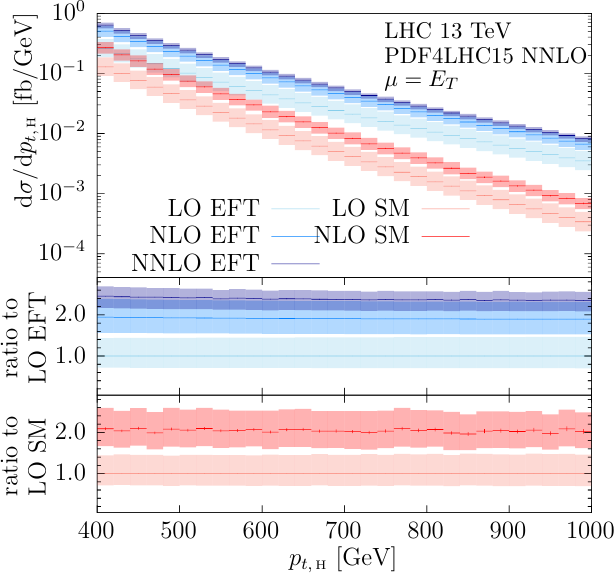}
\caption{Transverse momentum distribution of the Higgs boson at the
  LHC with $\sqrt{s}=13$~TeV computed in
  refs.~\cite{Chen:2016zka,Jones:2018hbb,Chen:2021azt}.  The upper
  panel shows absolute predictions at LO (${\cal O}(\alpha_s^3)$) and
  NLO (${\cal O}(\alpha_s^4)$) in the full SM and in the infinite
  $\mtop$ approximation (EFT), as well as the NNLO (${\cal
    O}(\alpha_s^5)$) in the EFT. The lower panels show the ratio of
  the EFT and full SM predictions to their respective LO
  calculations. The bands indicate theoretical errors obtained with a
  7-point scale variation, i.e. we perform a variation of $\mu_R$ and
  $\mu_F$ by a factor of two around their central value by keeping $
  1/2\leq \mu_R/\mu_F \leq 2$.}
\label{fig:fullNLO}
\end{figure*}

One way to improve on the pure EFT computation is to create
the so-called Born-improved EFT approximation.  To this end the EFT
cross section is simply rescaled by the exact leading order SM cross
section~\cite{Baur:1989cm,Ellis:1987xu}. For the inclusive
(cumulative) cross section, defined as
\begin{equation}
\Sigma(p_{\perp}^{\rm cut})=\int_{p_{\perp}^{\rm cut}}^\infty \frac{d\sigma}{dp_\perp^\prime} dp_\perp^\prime\,,
\end{equation}
this amounts to defining
\begin{equation}
\label{eq:rescaling_1}
\Sigma^{\text{EFT-improved (0), NNLO}}(p_{\perp}^{\rm cut})\equiv \frac{\Sigma^{\text{SM,
      LO}}(p_{\perp}^{\rm cut})}{\Sigma^{\text{EFT, LO}}(p_{\perp}^{\rm cut})} \,\Sigma^{\text{EFT, NNLO}}(p_{\perp}^{\rm cut})\,.
\end{equation}
The numerical implications of this Born-improved NNLO predictions were
first studied in ref.~\cite{Chen:2016zka} and show deviations from the
pure EFT computation at the level of $50 \%$ for transverse momenta of
$400$ GeV.  Since this modification is performed at leading order, a
considerable perturbative uncertainty has to be associated with this
procedure and higher order corrections are desirable.  In order to
further improve the result several approximations were considered in
refs.~\cite{Buschmann:2014sia,Frederix:2016cnl,Hamilton:2015nsa,Neumann:2016dny}
including exact real matrix elements at NLO in QCD and approximations
for virtual matrix elements. Finally, the two-loop virtual matrix
elements were included through an asymptotic expansion in
refs.~\cite{Lindert:2018iug,Neumann:2018bsx}, and exactly in
refs.~\cite{Jones:2018hbb,Chen:2021azt,Bonciani:2022jmb}, hence allowing for the
computation of the full NLO corrections.
The exact NLO QCD corrections computed in
refs.~\cite{Jones:2018hbb,Chen:2021azt,Bonciani:2022jmb} modify the exact leading
order prediction significantly but in a uniform way for the dynamical
scale chosen here, as it can be appreciated from
Fig.~\ref{fig:fullNLO}, from which one can observe a $K$ factor with a
very mild $p_\perp$ dependence.
An analogous behaviour is observed in the predictions obtained within
the EFT.  As a consequence, the modifications of the shape of the
$p_\perp$ distribution of the Higgs boson due to finite $\mtop$
effects is to a good extent already accounted for in
Eq.~\eqref{eq:rescaling_1} by the inclusion of exact leading order
matrix elements.
We collect in Table~\ref{tab:results_gionata} the inclusive cross
section $\Sigma$ for some relevant $p_\perp$ cuts up to both NNLO in
the EFT~\cite{Chen:2016zka} and to NLO in the full
SM~\cite{Jones:2018hbb,Chen:2021azt}. We will adopt the predictions from these
two references in the following study.

\begin{table*}[htp!]
\begin{adjustwidth}{-0.2cm}{}
\begin{tabular}{c|ccc||ccccc}
\multicolumn{1}{c}{} & \multicolumn{7}{c}{\large \hspace{2em} Inclusive cross
                       sections ([fb]) and $K$-factors for
  $pp\to H+X$}\\\hline\hline
 $p_T^\mathrm{cut}$ & $\mathrm{LO}_\mathrm{full}$ & $\mathrm{NLO}_\mathrm{full}$ & $K_\mathrm{full}^\mathrm{NLO}$  &  $\mathrm{LO}_\mathrm{EFT}$ & $\mathrm{NLO}_\mathrm{EFT}$ & $\mathrm{NNLO}_\mathrm{EFT}$ & $K_\mathrm{EFT}^\mathrm{NLO}$  & $K_\mathrm{EFT}^\mathrm{NNLO}$ \\ \hline
400 & $11.9^{+45\%}_{-29\%}$ & $24^{+24\%}_{-20\%}$ & 2.06 & $32^{+44\%}_{-29\%}$ & $63^{+23\%}_{-19\%}$ & $78^{+9.2\%}_{-12\%}$ & 1.93 & 1.25 \\
450 & $6.5^{+45\%}_{-29\%}$ & $13.3^{+24\%}_{-20\%}$ & 2.05 & $21^{+45\%}_{-29\%}$ & $41^{+22\%}_{-19\%}$ & $51^{+8.9\%}_{-11\%}$ & 1.92 & 1.25 \\
500 & $3.7^{+45\%}_{-29\%}$ & $7.5^{+24\%}_{-20\%}$ & 2.05 & $14.2^{+45\%}_{-29\%}$ & $27^{+22\%}_{-20\%}$ & $34^{+8.8\%}_{-11\%}$ & 1.91 & 1.25 \\
550 & $2.1^{+45\%}_{-30\%}$ & $4.4^{+24\%}_{-20\%}$ & 2.04 & $9.8^{+45\%}_{-29\%}$ & $18.6^{+22\%}_{-20\%}$ & $23^{+8.8\%}_{-11\%}$ & 1.91 & 1.25 \\
600 & $1.28^{+46\%}_{-30\%}$ & $2.6^{+24\%}_{-20\%}$ & 2.03 & $6.8^{+45\%}_{-29\%}$ & $13.0^{+22\%}_{-20\%}$ & $16.2^{+8.8\%}_{-11\%}$ & 1.90 & 1.24 \\
650 & $0.79^{+46\%}_{-30\%}$ & $1.60^{+24\%}_{-20\%}$ & 2.03 & $4.9^{+46\%}_{-29\%}$ & $9.3^{+22\%}_{-20\%}$ & $11.5^{+8.7\%}_{-11\%}$ & 1.90 & 1.24 \\
700 & $0.49^{+47\%}_{-30\%}$ & $1.00^{+24\%}_{-20\%}$ & 2.03 & $3.5^{+46\%}_{-29\%}$ & $6.7^{+22\%}_{-20\%}$ & $8.3^{+8.7\%}_{-11\%}$ & 1.90 & 1.24 \\
750 & $0.32^{+47\%}_{-30\%}$ & $0.64^{+24\%}_{-20\%}$ & 2.03 & $2.6^{+46\%}_{-30\%}$ & $4.9^{+22\%}_{-20\%}$ & $6.1^{+8.7\%}_{-11\%}$ & 1.90 & 1.24 \\
800 & $0.20^{+47\%}_{-30\%}$ & $0.41^{+24\%}_{-20\%}$ & 2.01 & $1.90^{+46\%}_{-30\%}$ & $3.6^{+22\%}_{-20\%}$ & $4.5^{+8.7\%}_{-11\%}$ & 1.90 & 1.24 \\
850 & $0.135^{+47\%}_{-30\%}$ & $0.27^{+24\%}_{-20\%}$ & 2.00 & $1.42^{+47\%}_{-30\%}$ & $2.7^{+22\%}_{-20\%}$ & $3.3^{+8.7\%}_{-11\%}$ & 1.89 & 1.24 \\
900 & $0.090^{+47\%}_{-30\%}$ & $0.180^{+23\%}_{-20\%}$ & 2.00 & $1.07^{+47\%}_{-30\%}$ & $2.0^{+22\%}_{-20\%}$ & $2.5^{+8.5\%}_{-11\%}$ & 1.89 & 1.24 \\
950 & $0.061^{+48\%}_{-30\%}$ & $0.120^{+23\%}_{-20\%}$ & 1.98 & $0.81^{+47\%}_{-30\%}$ & $1.53^{+22\%}_{-20\%}$ & $1.90^{+8.6\%}_{-11\%}$ & 1.89 & 1.24 \\
1000 & $0.041^{+48\%}_{-30\%}$ & $0.081^{+24\%}_{-20\%}$ & 1.95 & $0.62^{+47\%}_{-30\%}$ & $1.17^{+22\%}_{-20\%}$ & $1.45^{+8.6\%}_{-11\%}$ & 1.89 & 1.24 \\
1050 & $0.029^{+48\%}_{-30\%}$ & $0.056^{+23\%}_{-20\%}$ & 1.96 & $0.47^{+47\%}_{-30\%}$ & $0.90^{+22\%}_{-20\%}$ & $1.12^{+8.6\%}_{-11\%}$ & 1.89 & 1.24 \\
1100 & $0.0199^{+49\%}_{-30\%}$ & $0.039^{+24\%}_{-20\%}$ & 1.94 & $0.37^{+48\%}_{-30\%}$ & $0.69^{+22\%}_{-20\%}$ & $0.86^{+8.7\%}_{-11\%}$ & 1.89 & 1.24 \\
1150 & $0.0139^{+49\%}_{-30\%}$ & $0.027^{+24\%}_{-20\%}$ & 1.92 & $0.28^{+48\%}_{-30\%}$ & $0.54^{+22\%}_{-20\%}$ & $0.67^{+8.7\%}_{-11\%}$ & 1.90 & 1.24 \\
1200 & $0.0098^{+49\%}_{-31\%}$ & $0.0186^{+24\%}_{-20\%}$ & 1.90 & $0.22^{+48\%}_{-30\%}$ & $0.42^{+22\%}_{-20\%}$ & $0.52^{+8.7\%}_{-12\%}$ & 1.90 & 1.24 \\
1250 & $0.0070^{+49\%}_{-31\%}$ & $0.0130^{+25\%}_{-20\%}$ & 1.86 & $0.173^{+48\%}_{-31\%}$ & $0.33^{+22\%}_{-20\%}$ & $0.41^{+8.6\%}_{-12\%}$ & 1.90 & 1.24 \\
\end{tabular}
\caption{ Inclusive cross sections in fb and $K$-factors for
  $pp\to H+X$ in the SM for the relevant $p^{\rm cut}_\perp$ values
  (in GeV units) as computed in
  refs.~\cite{Chen:2016zka,Jones:2018hbb,Chen:2021azt}. Uncertainties are estimated
  by varying $\mu_F$ and $\mu_R$ separately by factors of $1/2$ and
  $2$ while keeping $ 1/2\leq \mu_R/\mu_F \leq 2$.  The $K$-factors
  are defined as
  $K_\mathrm{SM}^\mathrm{NLO} = \mathrm{NLO}_\mathrm{SM} /
  \mathrm{LO}_\mathrm{SM}$,
  $K_\mathrm{EFT}^\mathrm{NLO} = \mathrm{NLO}_\mathrm{EFT} /
  \mathrm{LO}_\mathrm{EFT}$,
  and
  $K_\mathrm{EFT}^\mathrm{NNLO} = \mathrm{NNLO}_\mathrm{EFT} /
  \mathrm{NLO}_\mathrm{EFT}$.  }
\label{tab:results_gionata}
\end{adjustwidth}
\end{table*}

Ideally, we want to combine the NNLO predictions computed in the EFT
with the exact NLO prediction. Under the assumption that the exact
NNLO QCD corrections follow the pattern of the NNLO EFT corrections,
i.e. they would lead to a uniform K-factor, this can be achieved by
rescaling EFT NNLO predictions in the following way:
\begin{equation}
\label{eq:approx_NNLO}
\Sigma^{\text{EFT-improved (1), NNLO}}(p_{\perp}^{\rm cut})\equiv \frac{\Sigma^{\text{SM,
      NLO}}(p_{\perp}^{\rm cut})}{\Sigma^{\text{EFT, NLO}}(p_{\perp}^{\rm cut})} \,\Sigma^{\text{EFT, NNLO}}(p_{\perp}^{\rm cut})\,.
\end{equation}

\begin{figure}[htb]
\centering
\includegraphics[width=0.9\textwidth]{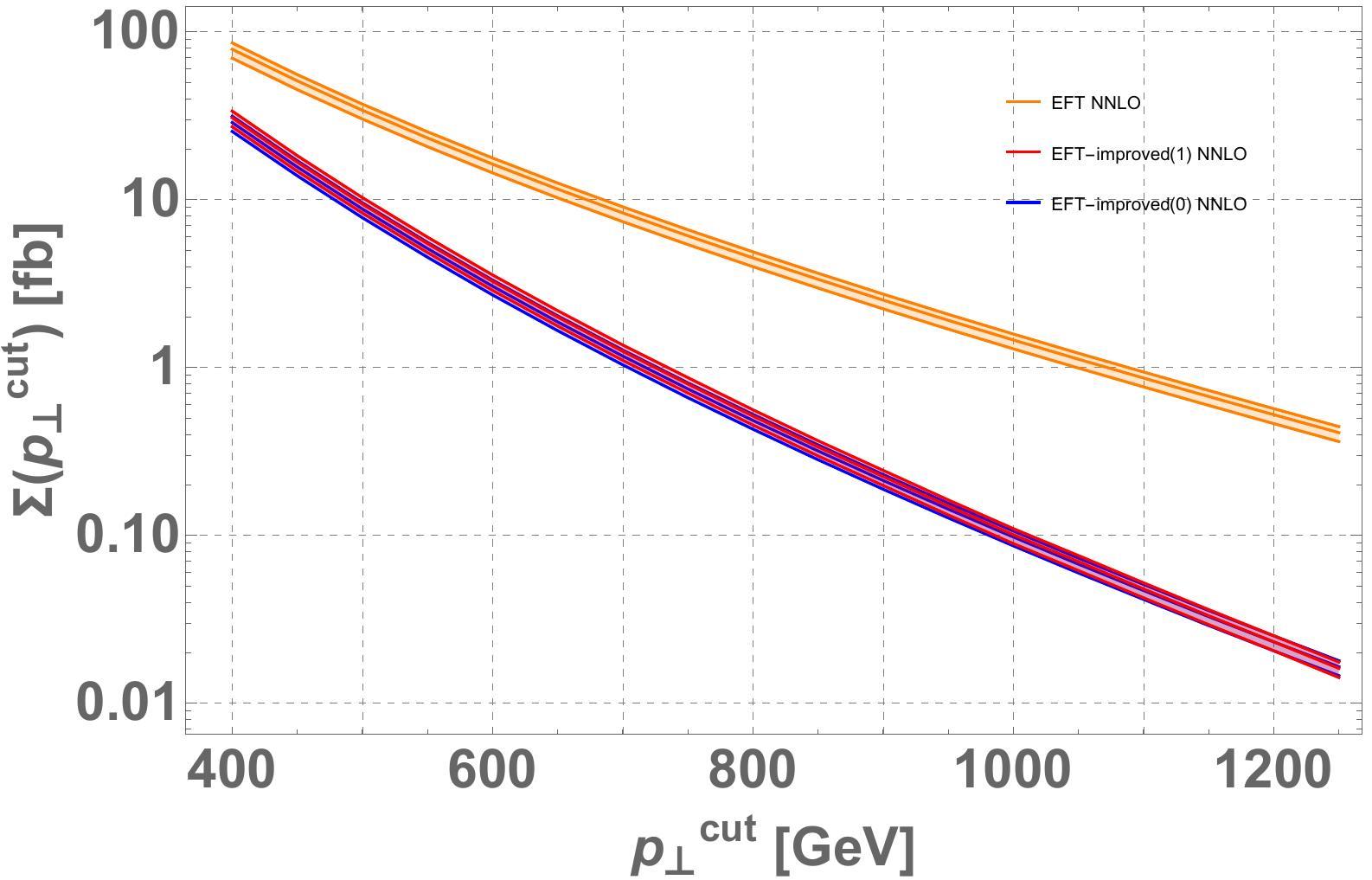}
\caption{Cumulative cross section
as a function of the $p_\perp$ cut at NNLO in the heavy-top EFT, as well as rescaled by the LO (NLO) full-SM
spectrum labelled by EFT-improved$(0)$ (EFT-improved$(1)$). See the text for description. The
ratio of the EFT-improved$(1)$ and EFT-improved$(0)$ predictions is
shown in Fig.~\ref{fig:relativefinal}
\label{fig:absolutfinal}
}
\end{figure}

\begin{figure}[htb]
\hspace{0.5cm}\includegraphics[width=0.9\textwidth]{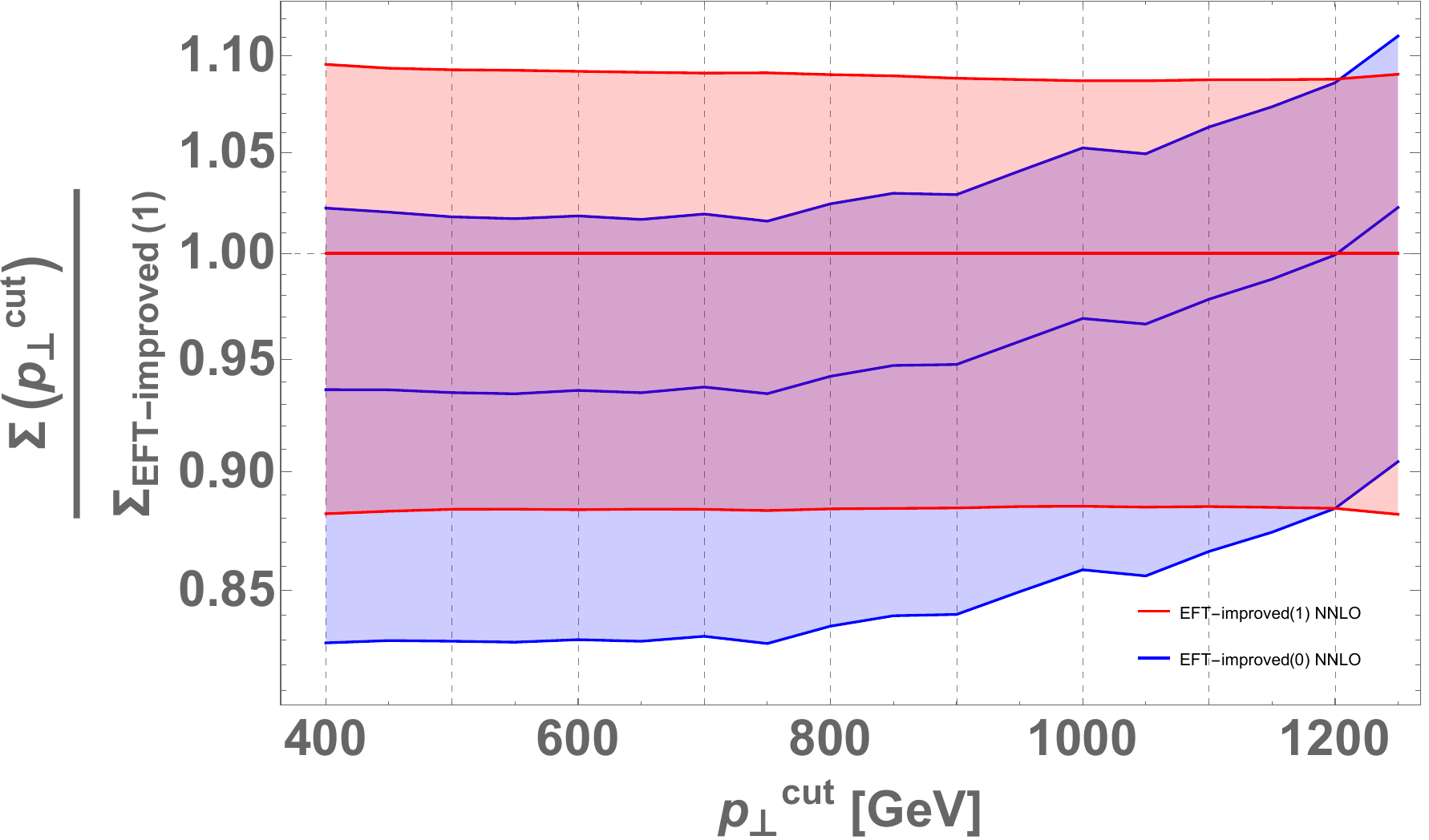}
\caption{Ratio of the cumulative cross section as defined in the
  EFT-improved$(0)$ and EFT-improved$(1)$ approximation (see the text for description) to the central value of
  the EFT-improved$(1)$ result
as a function of the $p_\perp$ cut.
\label{fig:relativefinal}
}
\end{figure}

We quote the prediction obtained with Eq.~\eqref{eq:approx_NNLO} as
the current best prediction.\footnote{We point out that the rescaling
  performed in Eqs.~\eqref{eq:rescaling_1},~\eqref{eq:approx_NNLO}
  could be alternatively defined at the differential level,
  leading to yet another prescription to combine consistently the NNLO
  prediction in the EFT with the NLO calculation in the full SM. Since
  in this document we will only refer to the cross section
  $\Sigma(p_\perp^{\rm cut})$ we choose to perform the rescaling at
  the level of the cumulative cross section.} To estimate the theory
uncertainty in the resulting cross section we proceed as follows:
\begin{itemize}
\item We perform a variation of $\mu_R$ and $\mu_F$ by a factor of two
  around their central value by keeping $ 1/2\leq \mu_R/\mu_F \leq 2$
  (7 point scale variation). The scales are varied separately in
  $\Sigma^{\text{EFT, NNLO}} $ and in the
  $\Sigma^{\text{SM, NLO}}/\Sigma^{\text{EFT, NLO}}$ ratio. For the
  latter, the same scale is chosen for the numerator and the
  denominator, and the final uncertainty is symmetrised. Finally, the
  two uncertainties are combined either in quadrature or linearly.
\item We assume that the uncertainty due to mass effects in the NNLO
  EFT correction is obtained by rescaling the latter by the relative
  mass correction at NLO. Thus, we assess the uncertainty
  $\delta_{\text{NNLO},\,m_t}$ as
\begin{eqnarray}
\delta_{\text{NNLO},\,m_t}&\equiv& \frac{\delta \Sigma^{\text{SM,
                                   NLO}}- \delta
                                   \Sigma^{\text{improved(0),
                                   NLO}}}{\delta \Sigma^{\text{EFT, NLO}}}
                                   \times \delta  \Sigma^{\text{EFT,
                                   NNLO}}\nonumber\\
&=& \frac{\delta \Sigma^{\text{SM, NLO}}- \delta \Sigma^{\text{improved(0), NLO}}}{\delta \Sigma^{\text{improved(0), NLO}}} \times \delta  \Sigma^{\text{improved(0), NNLO}}.
\end{eqnarray}
Here, $\delta\Sigma$ refers to the perturbative correction at a given order in QCD perturbation theory,
namely $\delta  \Sigma^{\text{X, (N)NLO}} = \Sigma^{\text{X, (N)NLO}} - \Sigma^{\text{X, (N)LO}}$.
\item The final uncertainty is obtained by combining the scale and
  mass effect uncertainties defined in the previous two items. In
  Table~\ref{tab:FOresults} we report the results for the cross
  sections, where the uncertainties are either combined in quadrature
  (NNLO$^{\rm approximate}_{\rm quad. unc.}$) or summed linearly
  (NNLO$^{\rm approximate}_{\rm lin. unc.}$). In the following, we
  work under the assumption that the three sources of uncertainty are
  uncorrelated, and therefore will consider the combination in
  quadrature as our central prescription.
\item An additional source of uncertainty is given by the top-mass
  scheme, for which we adopt the on-shell scheme used in the
  calculation of refs.~\cite{Jones:2018hbb,Chen:2021azt}. The
  difference between the on-shell and the ${\rm \overline{MS}}$ scheme
  has been recently studied in
  ref.~\cite{Bonciani:2022jmb}\footnote{The top mass scheme was also
  discussed in Ref.~\cite{Amoroso:2020lgh}.}, and shown to be
  substantial at LO for typical renormalisation scales in boosted
  Higgs production. In the same reference it was shown that
  differences are reduced at NLO, although they are still substantial
  and warrant careful consideration. We leave this discussion for
  future work of the working group.

\end{itemize}
{\renewcommand{\arraystretch}{1.5}
\begin{table*}[htp!]
 \vspace*{0.3ex}
 \begin{center}
\begin{small}
   \begin{tabular}{c|c|c}
$p_\perp^{\rm cut}$& 
 
  NNLO$^{\rm approximate}_{\rm quad. unc.}$  [fb] & NNLO$^{\rm
                                                    approximate}_{\rm
                                                    lin. unc.}$  [fb] \\
\hline
$400~\text{GeV}$    & $30.7^{+9.6\%}_{-11.8\%}$    & $30.7^{+11.9\%}_{-14.2\%}$             
\\ 
$430~\text{GeV}$  & $21.2^{+9.6\%}_{-11.8\%}$     & $21.2^{+11.9\%}_{-14.2\%}$
\\ 
$450~\text{GeV}$  & $16.7 ^{+9.5\%}_{-11.8\%}$   & $16.7 ^{+11.9\%}_{-14.2\%}$
   \end{tabular}
\end{small}
 \end{center}
 \caption{Best prediction $\Sigma^{\text{EFT-improved (1), NNLO}}$ for
   the inclusive cross sections at different $p_\perp$ cuts of
   phenomenological interest, and using two different prescriptions for
   the uncertainty (see text for details). 
 }
\label{tab:FOresults}
\end{table*}
}In Fig.~\ref{fig:absolutfinal} we show the cumulative cross section
as a function of the $p_\perp$ cut. The figure compares the NNLO EFT,
Born-improved NNLO EFT (EFT-improved$(0)$) and our best prediction
(EFT-improved$(1)$), obtained using
Eq.~\eqref{eq:approx_NNLO}. Fig.~\ref{fig:relativefinal} shows the
ratio of the latter two predictions to the central value of the
EFT-improved$(1)$ prediction. The uncertainties in the
EFT-improved$(0)$ band has been obtained by pure scale variation,
while the uncertainty in the EFT-improved$(1)$ prediction is estimated
as outlined above.

%% file: Chapters/MC.tex
\subsection{Event generators}
\label{sec:generators}
In this section we report the predictions obtained with different 
event generators for the boosted-Higgs scenario.

We compare the following Monte-Carlo tools:
\begin{itemize}
\item {\tt POWHEG gg$_{-}$h}~\cite{Alioli:2008tz}: NLO accurate for
  inclusive gluon fusion and LO (${\cal O}(\alpha_s^3)$) in the
  $p_\perp$ spectrum. The calculation is performed in the heavy-top
  EFT. The default {\tt POWHEG} $\mu_R$ and $\mu_F$ scales are
  used. The {\tt hfact} parameter~\cite{Alioli:2008tz} is set to
  $h=104$~GeV as in the CMS analysis~\cite{Sirunyan:2017dgc} (this
  only impacts the predictions matched to a parton shower below).
\item {\tt POWHEG HJ}~\cite{Campbell:2012am}: NLO accurate (${\cal O}(\alpha_s^4)$) in the
  Higgs $p_\perp$ spectrum. The calculation is performed in the heavy-top
  EFT. $\mu_R$ and $\mu_F$ are set to
  $H_T/2=1/2\left(\sqrt{m_H^2+p_\perp^2} +
    \sum_{i=1}^n|p_{t,i}|\right)$,
  where $p_{t,i}$ is the transverse momentum of the $i$-th radiated
  parton (n = 1 for Born/Virtual events, n = 2 for real events).
\item {\tt HJ-MiNLO}~\cite{Hamilton:2012rf}: NLO for inclusive gluon
  fusion and NLO in the $p_\perp$ spectrum. $\mu_R$ and $\mu_F$ are
  always set to $p_\perp$. The calculation is performed in the
  heavy-top EFT, but finite $\mtop$ effects can be included via a
  rescaling by the LO spectrum in the full SM. Born events
  with one jet terms are proportional to
  $\alpha_s^2(m_H)\alpha_s(p_\perp)$, while NLO corrections are
  proportional to $\alpha_s^2(m_H)\alpha^2_s(p_\perp)$.
\item {\tt MG5$_{-}$MC@NLO}~\cite{Alwall:2014hca,Frederix:2016cnl}:
  predictions obtained by merging samples of $0$,$1$, and $2$ jets,
  NLO accurate for all the above multiplicities.  Finite $\mtop$
  corrections are included exactly in the Born and real corrections
  for all multiplicities, and approximately in the virtual corrections
  by rescaling the heavy-top EFT virtual corrections by the LO result
  in the full SM.  The scale is set following the {\tt
    FxFx}~\cite{Frederix:2012ps} prescription and the merging scale is
  set to $30$~GeV. The merging scale sets the effective momentum scale
  at which the event sample transitions between the various
  jet-multiplicities, cf. Section 2 of Ref.~\cite{Frederix:2012ps}.
\end{itemize}

The results for the {\tt POWHEG}/{\tt MiNLO} generators are reported
both at fixed order and matched to the {\tt Pythia 6} parton shower
Monte Carlo~\cite{Sjostrand:2006za}, in Table~\ref{tab:MC_FO}
and~\ref{tab:MC_PS}, respectively. Table~\ref{tab:MC_FO} shows the
predictions from the {\tt POWHEG}/{\tt MiNLO} generators before the
matching to a parton shower is performed, while Table~\ref{tab:MC_PS}
reports the predictions matched to a parton shower simulation.
The last row of the tables shows the result of {\tt HJ-MiNLO}
including mass effects, as implemented in
ref.~\cite{Hamilton:2015nsa}. The results include only the top
contribution, implemented through a rescaling of the EFT result by the
exact LO spectrum, and hence very similar in spirit to the
prescription introduced in Section~\ref{sec:fixed}, in
Eq.~\eqref{eq:rescaling_1}. In the large Higgs transverse momentum
region, the generator {\tt HJ-MiNLO} reproduces exactly the {\tt
  NNLOPS}~\cite{Hamilton:2013fea} and {\tt MiNNLO$_{\rm PS}$}
~\cite{Monni:2019whf,Monni:2020nks} generators currently used in Higgs
analyses for the gluon fusion channel at the LHC. Uncertainties are
obtained through a $7$-point scale variation around the central
renormalisation and factorisation scales by a factor of two.


{\renewcommand{\arraystretch}{1.5}
\begin{table*}[htp!]
  \vspace*{0.3ex}
  \begin{center}
\begin{small}
    \begin{tabular}{||c||c|c|c|c||}
\hline\hline
Fixed order level [pb] & Total & $p_\perp^{\rm cut} > 400$ GeV & $p_\perp^{\rm
                                                        cut} > 450$
                                                            GeV &
                                                                  $p_\perp^{\rm cut} > 500$ GeV\\
\hline\hline
{\tt gg$_{}$h$^{\rm hfact=104}_{\rm m_t=\infty}$} & $30.3$ & $0.0730$&$0.0507$ &$0.0362$\\
\hline
{\tt HJ} $m_t=\infty$, 5 GeV gen. cut & $-$ & $0.0643$&$0.0413$ &$0.0278$ \\
\hline
{\tt HJ} $m_t=\infty$, 50 GeV gen. cut & $-$ & $0.0644$& $0.0416$& $0.0277$\\
\hline
{\tt HJ-MiNLO} $m_t=\infty$ & $32.1$ &$0.0778$ &$0.0509$ &$0.0343$ \\
\hline
{\tt HJ-MiNLO} $m_t=171.3$ GeV & $33.8$ &$0.0281$ &$0.0153$ &$0.0089$ \\
\hline\hline
    \end{tabular}
\end{small}
  \end{center}
\caption{Results from the indicated event generators for different
  $p_\perp$ cuts (in GeV units) before the matching to parton showers
  is performed (labelled as {\it Fixed order level} in the
  table). Predictions are expressed in [pb] units. The total cross
  section for $gg\to H$ obtained with the indicated event generator is
  also reported whenever available. The total cross section shown here
  is by construction identical to the one reported in
  Table~\ref{tab:MC_PS} - including the scale uncertainties.}
\label{tab:MC_FO}
\end{table*}
}


{\renewcommand{\arraystretch}{1.5}
\begin{table*}[htp!]
  \vspace*{0.3ex}
  \begin{center}
\begin{small}
    \begin{tabular}{||c||c|c|c|c||}
\hline\hline
Parton shower matched level [pb] & Total & $p_\perp^{\rm cut} > 400$ GeV & $p_\perp^{\rm
                                                        cut} > 450$
                                                            GeV &
                                                                  $p_\perp^{\rm cut} > 500$ GeV\\
\hline\hline
{\tt gg$_{}$h$^{\rm hfact=104}_{\rm m_t=\infty}$} & $30.3^{+6.1}_{-4.7}$ & $0.0829^{+0.0451}_{-0.0266}$&$0.0577^{+0.0325}_{-0.019}$ &$0.0408^{+0.0236}_{-0.0137}$\\
\hline
{\tt HJ} $m_t=\infty$, 5 GeV gen. cut & $-$ & $0.0651^{+0.0156}_{-0.0131}$&$0.0417^{+0.01}_{-0.0084}$ &$0.0279^{+0.0067}_{-0.0057}$ \\
\hline
{\tt HJ} $m_t=\infty$, 50 GeV gen. cut & $-$ & $0.0651^{+0.0156}_{-0.0131}$& $0.0418^{+0.01}_{-0.0085}$& $0.0278^{+0.0066}_{-0.0056}$ \\
\hline
{\tt HJ-MiNLO} $m_t=\infty$ & $32.1^{+11}_{-4.9}$ &$0.0803^{+0.9087}_{-0.0164}$ &$0.0524^{+0.0118}_{-0.0107}$ &$0.0353^{+0.0078}_{-0.0072}$ \\
\hline
{\tt HJ-MiNLO} $m_t=171.3$ GeV & $33.8^{+11.4}_{-5.2}$ &$0.029^{+0.007}_{-0.006}$ &$0.0161^{+0.0036}_{-0.0033}$ &$0.0091^{+0.0021}_{-0.0018}$ \\
\hline\hline
    \end{tabular}
\end{small}
  \end{center}
\caption{Results matched to parton shower for different $p_\perp$ cuts (in GeV units) for the indicated event
  generators. Predictions are expressed in [pb] units. The total cross
  section for $gg\to H$ obtained with the indicated event generator is also reported whenever available.}
\label{tab:MC_PS}
\end{table*}
}

By inspecting the last two rows of Tables~\ref{tab:MC_FO}
and~\ref{tab:MC_PS}, we observe that the inclusion of the parton
shower has a moderate impact on the result (at the $2-5\%$ level), as
one expects for the considered kinematics regime.

The results obtained with {\tt MG5$_{-}$MC@NLO} are obtained with top
mass corrections included exactly in the Born and real corrections,
and approximately in the virtual corrections by rescaling the EFT
virtual corrections by the LO result in the full SM. Exact bottom
quark mass effects are not included as they are negligible in the
considered region. The events are showered with the {\tt Pythia 8}
parton shower Monte Carlo~\cite{Sjostrand:2014zea}. The results for
some relevant $p_\perp$ cuts are summarised in
Table~\ref{tab:MCNLOresults}, together with a comparison to the
results of the {\tt HJ-MiNLO} generator, and to our best prediction
described in Section~\ref{sec:fixed}. The quoted uncertainties have
been obtained by a $9$-point scale variation, i.e. independently
around the central renormalisation and factorisation scales by a
factor of two.\footnote{The uncertainty prescriptions adopted in
Table~\ref{tab:MCNLOresults} reflect the nominal prescriptions used in
Refs.~\cite{Hamilton:2012rf,Frederix:2016cnl}. For the fixed order
prediction we adopt the prescription discussed in the
Section~\ref{sec:fixed}.}

{\renewcommand{\arraystretch}{1.5}
\begin{table*}[htp!]
  \vspace*{0.3ex}
  \begin{center}
\begin{small}
    \begin{tabular}{c|c|c|c}
$p_\perp^{\rm cut}$ & 
  
   NNLO$^{\rm approximate}_{\rm quad. unc.}$ [fb]  & 
   {\tt HJ-MINLO} [fb]  &
   {\tt MG5\_MC@NLO} [fb]    \\
\hline
$400~\text{GeV}$   & $30.7^{+9.6\%}_{-11.8\%}$    &   $29^{+24\%}_{-21\%} $    & $31.5^{+31\%}_{-25\%}$ 
\\ 
$450~\text{GeV}$  & $16.7 ^{+9.5\%}_{-11.8\%}$   &    $16.1^{+22\%}_{-21\%} $    &  $17.1 ^{+31\%}_{-25\%}$ 
    \end{tabular}
\end{small}
  \end{center}
  \caption{Comparison of predictions at fixed order in the
    $\Sigma^{\text{EFT-improved (1), NNLO}}$ approximation, with {\tt
      HJ-MINLO}  and with {\tt MG5$_{-}$MC@NLO}. The uncertainties in
    the three predictions are
    obtained by means of a 7 points scale variation (NNLO$^{\rm
      approximate}_{\rm quad. unc.}$ and {\tt HJ-MINLO}), and 9 point scale variation ({\tt MG5\_MC@NLO}),
    respectively. The difference in the uncertainty prescription is
    reflected in the different theoretical errors quoted in the table.
    See text for more details.
  }
\label{tab:MCNLOresults}
\end{table*}
}

We observe that the predictions obtained with the more accurate
generators used in the study ({\tt HJ-MiNLO} and {\tt
  MG5$_{-}$MC@NLO}) are in very good agreement with one
another. Moreover, they both reproduce, within uncertainties, the best
prediction obtained in the previous section. We conclude that the
above two generators can be safely used to perform accurate studies in
the boosted regime. 
However, state of the art QCD predictions reach a higher level of
precision and novel methods are necessary to exploit such calculations
in the context of Monte Carlo simulations.

%% file: Chapters/Channels.tex
\section{Predictions for other production modes}
\label{app:channels}
In this Section we report the breakdown of the boosted Higgs cross
section into different production channels. 
In the following we consider both QCD and EW perturbative
corrections. We start by discussing the former, for which we consider
the same YR4 setup~\cite{YR4} discussed in Section~\ref{sec:fixed}
unless stated otherwise.
For vector boson fusion (VBF), the prediction is obtained from
refs.~\cite{Cacciari:2015jma}, where the VBF cross section is computed
to NNLO accuracy in perturbative QCD (${\cal O}(\alpha_s^2)$) obtained
in the so called {\it factorised} approximation~\cite{Han:1992hr}. In
the same approximation, N$^3$LO corrections are
known~\cite{Dreyer:2016oyx}, but are negligible for the accuracy
considered in this work. Non-factorising corrections have been
recently estimated~\cite{Liu:2019tuy,Dreyer:2020urf}, and it was
concluded that they may be potentially relevant in the considered
phase space region. Nevertheless, we do not expect these corrections
to affect our qualitative conclusions. For this process we set the
renormalisation and factorisation scales to
$\mu_R^2 = \mu_F^2 = m_H/2 \sqrt{(m_H/2)^2+p_{\perp}^2}$. Perturbative
uncertainties are obtained by varying both scales by a factor of two
while keeping $\mu_R = \mu_F$ ($3$-point variation).
For associated production VH ($V=W^{\pm}, Z$), we consider NLO (${\cal O}(\alpha_s)$)
predictions obtained with the {\tt
  POWHEG-BOX-V2}~\cite{Luisoni:2013kna,Alioli:2010xd}.\footnote{We
  note that, in addition, the $ZH$ channel may receive large
  perturbative correction to the gluon-induced subprocess.} 
The scales are set to the invariant mass of the $VH$ system as
$\mu_R = \mu_F = \sqrt{(p_H+p_V)^2}$, and perturbative uncertainties
are again obtained by varying both scales by a factor of two while
keeping $\mu_R = \mu_F$ ($3$-point variation).
Also in this case NNLO corrections are known to be small, with the
exception of the contribution from gluon fusion~\cite{YR4}. Therefore,
we do not include them in the following.
For VBF and VH we use a $3$- rather than $7$-point variation, because
the latter has been found to be almost entirely contained within the
former~\cite{Cacciari:2015jma,Astill:2016hpa}. We have explicitly
verified that this is the case in the boosted regime considered here.
Finally, for ${\rm t\bar t H}$, we consider NLO (${\cal
  O}(\alpha_s^3)$) predictions obtained
with {\tt
  Sherpa+OpenLoops}~\cite{Gleisberg:2008ta,Cascioli:2011va}. In this
case the perturbative scales are set to
$\mu_R = \mu_F = (M_{T,t}+M_{T,\bar t} + M_{T,H})/2$, and
uncertainties are obtained with a $7$-point variation.

The results are reported in Table~\ref{tab:channels}. We stress that
the quoted uncertainty only accounts for QCD scale variations
estimated as outlined above, and it does not contain PDF and
$\alpha_s$ errors.

For all channels but gluon fusion, NLO EW corrections have been known
for some time
(cf. ref.~\cite{Ciccolini:2003jy,Ciccolini:2007ec,Ciccolini:2007jr,Denner:2011id,Frixione:2014qaa,Yu:2014cka,Frixione:2015zaa,Denner:2016wet,YR4}),
and are obtained here using {\tt
  Sherpa+OpenLoops}~\cite{Gleisberg:2008ta,Cascioli:2011va,Kallweit:2014xda,Schonherr:2017qcj,Buccioni:2019sur}. The
emission of weak gauge bosons is not included in the EW corrections,
and should be considered as separate background reactions. We report
the results in Table~\ref{tab:EW}, which displays the percentage
decrease of the corresponding cross sections of
Table~\ref{tab:channels} due to the inclusion of electro-weak
corrections.
The calculation of the EW corrections in the gluon fusion channel has
recently been considered in
refs.~\cite{Bonetti:2020hqh,Becchetti:2020wof,Bonetti:2022lrk}. Although
a complete calculation in the regime considered in this work is not
yet available, we stress that these corrections are expected to be
sizeable at large transverse momentum, and must be estimated for an
accurate prediction of the gluon-fusion production rate.
Moreover, we observe that this observable receives substantial
contributions from other production modes, which therefore must be
taken into account together with the gluon-fusion channel in
experimental analyses.

\begin{figure*}[htp!]
\centering
\includegraphics[width=.45\textwidth]{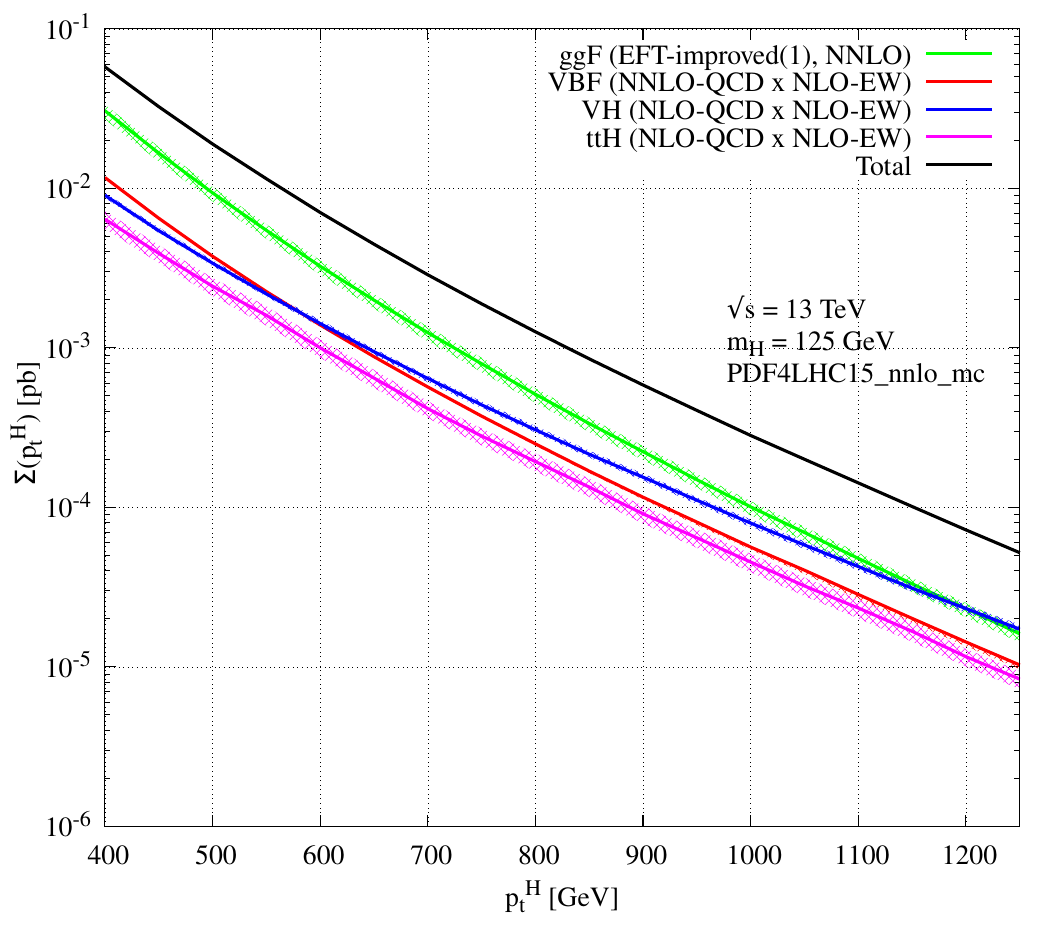}
\includegraphics[width=.45\textwidth]{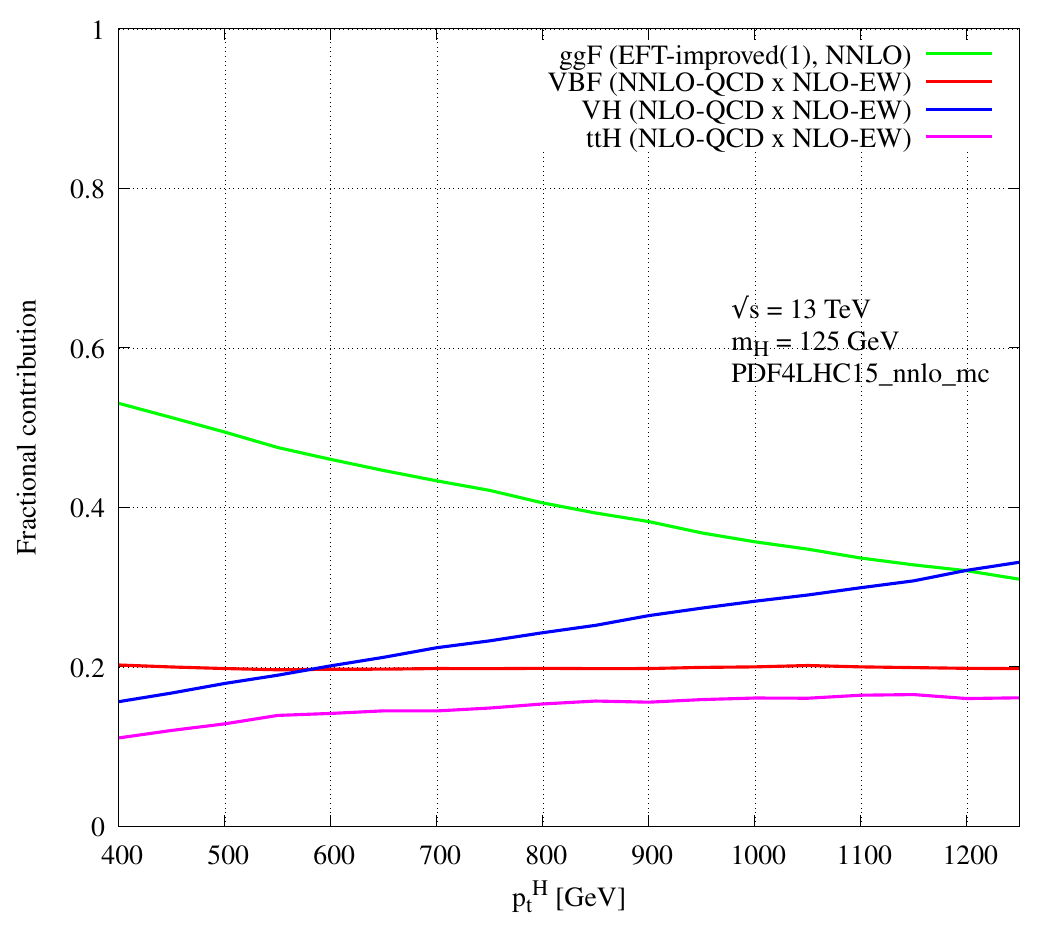}
\caption{ Cumulative cross section for the production of a Higgs boson
  as a function of the Higgs boson transverse momentum cut.  The
  cross section due to the gluon-fusion (green), VBF (red), vector
  boson associated (blue) and top-quark pair associated (magenta)
  production mode are shown in absolute values (left) and relative
  size (right).}
\label{fig:prodmodes}
\end{figure*}

Finally, the absolute and relative contributions of the different
production modes up to transverse momenta of $1.25$ TeV are summarised
in Fig.~\ref{fig:prodmodes}, including both QCD and EW corrections.
\begin{table*}[htp!]
\begin{equation}
\begin{array}{|c|c|c|c|c|}
\hline
p_\perp^{\rm cut} \text{[GeV]} & \Sigma^{\rm NNLO^{\rm approximate}_{\rm quad. unc.}}_{\rm ggF}( p_\perp^{\rm cut})
                                 \text{ [fb]} & \Sigma^{\rm NNLO}_{\rm
                                                VBF}(p_\perp^{\rm
                                                cut}) \text{ [fb]}  &
                                                                      \Sigma^{\rm NLO}_{\rm VH}(p_\perp^{\rm cut}) \text{ [fb]} &  \Sigma^{\rm NLO}_{\rm {\rm t\bar t H} }( p_\perp^{\rm cut})  \text{ [fb]} \\
\hline
400 & 30.67^{+9.59 \%}_{-11.84 \%}  & 14.23^{+0.15\%}_{-0.19\%} & 11.16^{+4.12\%}_{-3.68\%}& 6.89^{+12.62\%}_{-12.97\%} \\
 \hline
450 & 16.70^{+9.53 \%}_{-11.76 \%} & 8.06^{+0.24\%}_{-0.23\%} & 6.87^{+4.6\%}_{-3.49\%} & 4.24^{+12.84\%}_{-13.15\%} \\
 \hline
500 & 9.41^{+9.44 \%}_{-11.72 \%}& 4.75^{+0.33\%}_{-0.29\%} & 4.39^{+4.43\%}_{-4.04\%} & 2.66^{+12.85\%}_{-13.22\%} \\
 \hline
550 & 5.46^{+9.43 \%}_{-11.69 \%} & 2.90^{+0.34\%}_{-0.36\%} & 2.87^{+4.44\%}_{-3.74\%} & 1.76^{+14.23\%}_{-13.93\%} \\
 \hline
600 & 3.25^{+9.31 \%}_{-11.64 \%} & 1.82^{+0.41\%}_{-0.39\%} &  1.91^{+5.22\%}_{-4.71\%}  & 1.11^{+12.99\%}_{-13.4\%} \\
 \hline
650 & 1.99^{+9.21 \%}_{-11.63 \%}& 1.17^{+0.49\%}_{-0.39\%} &  1.30^{+4.67\%}_{-4.28\%}& 0.72^{+12.6\%}_{-13.26\%} \\
 \hline
700 & 1.24^{+9.09 \%}_{-11.57 \%}& 0.77^{+0.57\%}_{-0.45\%} &  0.90^{+4.15\%}_{-5.4\%} & 0.47^{+11.42\%}_{-12.74\%}\\
 \hline
750 & 0.79^{+9.16 \%}_{-11.60 \%}& 0.51^{+0.69\%}_{-0.56\%} & 0.62^{+5.15\%}_{-4.66\%} & 0.32^{+11.53\%}_{-12.84\%}\\
 \hline
800 & 0.51^{+9.05 \%}_{-11.56 \%}& 0.35^{+0.71\%}_{-0.6\%} & 0.44^{+5.64\%}_{-4.13\%} & 0.22^{+11.42\%}_{-13.3\%}\\
 \hline
 \end{array}
 \nonumber
\end{equation}
\caption{Predictions for the cumulative Higgs boson cross section as a
  function of the lower $p_\perp$ cut (the quoted gluon fusion cross section
  is obtained in the $\Sigma^{\text{EFT-improved (1), NNLO}}$ approximation). 
  We show QCD predictions for the various channels contributing to
  Higgs production. The table does not contain the EW corrections.
}
\label{tab:channels}
\end{table*}

\begin{table*}[htp!]
\begin{equation}
\begin{array}{|c|c|c|c|}
\hline
p_\perp^{\rm cut} \text{[GeV]} & {\rm VBF}  &{\rm VH} &  {\rm t\bar t H} \\
\hline
400 & -17.80\%& -19.05\%& -6.95\% \\
 \hline
450 & -19.43\%& -20.83\%& -7.75\%\\
 \hline
500 &  -21.05\% & -22.50\% & -8.49\%\\
 \hline
550 &  -22.34\%& -24.07\%& -9.11\%\\
 \hline
600 &  -23.73\%&  -25.56\%& -9.91\% \\
 \hline
650 &  -25.03\%&  -26.98\%& -10.67\%\\
 \hline
700 &  -26.29\% &  -28.30\%& -11.37\%\\
 \hline
750 &  -27.35\%& -29.60\%& -11.94\%\\
 \hline
800 &  -28.42\%& -30.83\% & -12.51\%\\
 \hline
 \end{array}
 \nonumber
\end{equation}
\caption{Percentage
  decrease of the cross sections of Table~\ref{tab:channels} due to the
  inclusion of electro-weak corrections as a
  function of the cut in $p_\perp$.
}
\label{tab:EW}
\end{table*}

%% file: Chapters/Conclusions.tex
\section{Summary and conclusions}
In this article we studied the inclusive production of a boosted Higgs
boson at the LHC. We presented a combination of accurate QCD
predictions for the various production channels, and provided a
recommendation for the cumulative distribution at large transverse
momenta in the gluon-fusion channel. The resulting predictions are
reported in Table~\ref{tab:channels} for different values of the lower
cut on the Higgs transverse momentum.
The table shows that in the boosted regime the dominance of the
gluon-fusion channel is much less significant, and a consistent
inclusion of different production modes is necessary. This is even
more important in view of BSM interpretations since different channels
can be affected differently by new-physics effects. It is therefore
desirable in experimental analyses to avoid subtracting different
Higgs production channels from the experimental measurement as a way
of assessing the gluon-fusion contribution. Such a subtraction can
only be done under strong theoretical assumptions. An unbiased way of
reporting the experimental results necessarily involves quoting the
fiducial cross sections.

For the gluon fusion contribution, we compare the resulting
predictions to those of Monte-Carlo event generators in
Table~\ref{tab:MCNLOresults} and find good agreement within the quoted
uncertainties. This implies that one can safely use the predictions
from the considered event generators with the associated theoretical
errors in the simulation of the boosted Higgs cross
section. Additional values of the gluon-fusion cross section are also
reported in Appendix~\ref{app:high-pT} up to scales of $1.25$ TeV.

We stress that we did not account here for other sources of
theoretical uncertainties (such as the top mass scheme, PDF and
couplings uncertainties, and EW corrections to the gluon-fusion
process), which must be included in the overall systematics in
phenomenological studies of the boosted Higgs cross section.

\paragraph{Acknowledgments} This work was done within the LHC Higgs Working Group (LHCHWG).
K.H. was supported by the Science and Technology Facilities Council
(STFC) under grant award ST/P000274/1, and by the European Commission
through the ERC Consolidator Grant HICCUP (No. 614577).

%% file: Chapters/gluon_fusion_high_pt.tex
\section{Gluon fusion cross section up to $1.25$ TeV}
\label{app:high-pT}
In this appendix we report additional predictions for the gluon fusion channel. Table~\ref{tab:finaltab} shows results in the range $p_\perp^{\rm cut}\in [400, 600]$ GeV for a finer binning than the one considered in the text, while cross section up to $p_\perp^{\rm cut}=1.25$ TeV in $50$-GeV bins are summarised in Table~\ref{tab:high-pT}.

\begin{table*}[htp!]
\begin{equation}
\begin{array}{|c|c|}
  \hline
  p_\perp^{\rm cut} \text{[GeV]} & \Sigma^{\rm NNLO^{\rm approximate}_{\rm quad. unc.}}_{\rm ggF}( p_\perp^{\rm cut}) \text{ [fb]} \\
  \hline
400 & 30.67^{+9.59 \%}_{-11.84 \%}  \\
 \hline
410 & 27.03^{+9.59 \%}_{-11.80 \%}  \\
 \hline
420 & 23.94^{+9.54 \%}_{-11.77 \%}  \\
 \hline
430 & 21.23^{+9.55 \%}_{-11.77 \%}  \\
 \hline
440 & 18.83^{+9.54 \%}_{-11.78 \%}  \\
 \hline
450 & 16.70^{+9.53 \%}_{-11.76 \%}  \\
 \hline
460 & 14.79^{+9.45 \%}_{-11.73 \%}  \\
 \hline
470 & 13.20^{+9.46 \%}_{-11.73 \%}  \\
 \hline
480 & 11.81^{+9.51 \%}_{-11.75 \%}  \\
 \hline
490 & 10.53^{+9.51 \%}_{-11.73 \%}  \\
 \hline
500 & 9.41^{+9.44 \%}_{-11.72 \%}  \\
 \hline
510 & 8.44^{+9.48 \%}_{-11.72 \%}  \\
 \hline
520 & 7.56^{+9.47 \%}_{-11.71 \%}  \\
 \hline
530 & 6.76^{+9.41 \%}_{-11.68 \%}  \\
 \hline
540 & 6.06^{+9.39 \%}_{-11.66 \%}  \\
 \hline
550 & 5.46^{+9.43 \%}_{-11.69 \%}  \\
 \hline
560 & 4.92^{+9.44 \%}_{-11.71 \%}  \\
 \hline
570 & 4.43^{+9.37 \%}_{-11.70 \%}  \\
 \hline
580 & 3.99^{+9.36 \%}_{-11.70 \%}  \\
 \hline
590 & 3.59^{+9.34 \%}_{-11.65 \%}  \\
 \hline
 \end{array}
 \nonumber
\end{equation}
\caption{Gluon fusion predictions for the cumulative Higgs boson cross section obtained in the $\Sigma^{\text{EFT-improved (1), NNLO}}$ approximation as a function of the lowest allowed $p_\perp$.   }
\label{tab:finaltab}
\end{table*}

\begin{table*}[htp!]
\begin{equation}
\begin{array}{|c|c|}
\hline
p_\perp^{\rm cut} \text{[GeV]} & \Sigma^{\rm NNLO^{\rm approximate}_{\rm quad. unc.}}_{\rm ggF}( p_\perp^{\rm cut})
                                 \text{ [fb]} \\
\hline
  400 & 30.67^{+9.59 \%}_{-11.84 \%}  \\
 \hline
450 & 16.70^{+9.53 \%}_{-11.76 \%}  \\
 \hline
500 & 9.41^{+9.44 \%}_{-11.72 \%}  \\
 \hline
550 & 5.46^{+9.43 \%}_{-11.69 \%}  \\
 \hline
600 & 3.25^{+9.31 \%}_{-11.64 \%}  \\
 \hline
650 & 1.99^{+9.21 \%}_{-11.63 \%}  \\
 \hline
700 & 1.24^{+9.09 \%}_{-11.57 \%}  \\
 \hline
750 & 0.80^{+9.16 \%}_{-11.60 \%}  \\
 \hline
800 & 0.51^{+9.05 \%}_{-11.56 \%}  \\
 \hline
850 & 0.34^{+8.93 \%}_{-11.58 \%}  \\
 \hline
900 & 0.22^{+8.81 \%}_{-11.56 \%}  \\
 \hline
950 & 0.15^{+8.74 \%}_{-11.50 \%}  \\
 \hline
1000 & 0.10^{+8.68 \%}_{-11.49 \%}  \\
 \hline
1050 & 0.07^{+8.68 \%}_{-11.53 \%}  \\
 \hline
1100 & 0.05^{+8.73 \%}_{-11.50 \%}  \\
 \hline
1150 & 0.03^{+8.73 \%}_{-11.54 \%}  \\
 \hline
1200 & 0.02^{+8.76 \%}_{-11.58 \%}  \\
 \hline
1250 & 0.02^{+9.02 \%}_{-11.84 \%}  \\
 \hline
 \end{array}
 \nonumber
\end{equation}
\caption{Gluon fusion cross section obtained in the $\Sigma^{\text{EFT-improved (1), NNLO}}$ approximation in highly boosted regime.}
\label{tab:high-pT}
\end{table*}